\documentclass[aps,prc,showpacs,floatfix]{revtex4}
\usepackage{graphicx}

\begin{document}

\draft
\title{The symmetry energy at subnuclear densities and nuclei in neutron star 
crusts}

\author{Kazuhiro Oyamatsu$^{1,2}$ and Kei Iida$^{2,3,4}$}
\affiliation{$^1$Department of Media Theories and Production, Aichi Shukutoku
University, Nagakute, Nagakute-cho, Aichi-gun, Aichi 480-1197, Japan\\
$^2$The Institute of Physical and Chemical Research (RIKEN),
Hirosawa, Wako, Saitama 351-0198, Japan\\
$^3$RIKEN BNL Research Center, Brookhaven National Laboratory,
Upton, NY 11973-5000, USA\\
$^4$Department of Materials Science, Kochi University,
Akebono-cho, Kochi 780-8520, Japan
}

\date{\today}

\begin{abstract}

     We examine how the properties of inhomogeneous nuclear matter at 
subnuclear densities depend on the density dependence of the symmetry energy.
Using a macroscopic nuclear model we calculate the size and shape of nuclei
in neutron star matter at zero temperature in a way dependent on the density 
dependence of the symmetry energy.  We find that for smaller symmetry energy 
at subnuclear densities, corresponding to larger density symmetry 
coefficient $L$, the charge number of nuclei is smaller, and the critical 
density at which matter with nuclei or bubbles becomes uniform is lower.  
The decrease in the charge number is associated with the dependence of the 
surface tension on the nuclear density and the density of a sea of neutrons,
while the decrease in the critical density can be generally 
understood in terms of proton clustering instability in uniform matter.

\end{abstract}
\maketitle

\section{Introduction}

     The inner crust of a neutron star consists of a Coulomb lattice of nuclei
embedded in a roughly uniform neutralizing background of electrons and in a
sea of neutrons \cite{PR}.  The equation of state (EOS) of uniform nuclear 
matter and the surface tension, both at large neutron excess, are essential to 
understanding of the equilibrium properties of matter in the crust.  However, 
laboratory data on nuclei only reflect the bulk and surface properties of 
nearly symmetric nuclear matter \cite{OI,IO}.  So far, calculations of the 
equilibrium properties of matter in the crust depend on the way to 
extrapolate the known bulk and surface properties to large neutron excess, 
which is different among earlier investigations \cite{PR}.

     In this paper we systematically analyze the question of how the 
equilibrium properties of inhomogeneous nuclear matter at subnuclear densities 
depend on the parameter characterizing the density dependence of the symmetry 
energy.  In doing so, we utilize a macroscopic nuclear model \cite{OI} in which
the equilibrium nucleon distribution depends on the EOS of nuclear matter 
through minimization of the energy density functional.  One of the most
important quantities is the charge number of the equilibrium nuclide.  In a 
liquid-drop picture \cite{PR}, this charge number is determined by the size 
equilibrium condition which controls the ratio between the Coulomb and 
surface energies.  This condition tells that 
the charge number squared is proportional to the surface tension
and the nuclear volume.  As we shall see, the density dependence of the 
symmetry energy, which controls the surface tension by affecting
the nuclear density and the density of the neutron gas, controls the charge 
number as well.

     We also address the question of how matter with nuclei 
or bubbles melts into uniform 
matter with increasing density.  In this melting process, rodlike and slablike
nuclei embedded in a gas of neutrons, as well as rodlike and roughly spherical
neutron-gas regions (bubbles) surrounded by a nucleon liquid, are expected to
occur \cite{RPW,HSY,LRP,O}.  At a density where roughly spherical nuclei are 
so closely packed that they occupy about 1/8 of the system volume, the nuclei 
tend to be elongated and eventually fuse into nuclear rods.  The advantage of
this rod formation is a reduction in the total surface area from the roughly 
spherical case.  However, whether bubbles and nonspherical nuclei actually 
appear in neutron star crusts depends on the critical density at which proton 
clustering instability occurs in uniform nuclear matter \cite{PRL}; they are 
expected to appear when the density corresponding to the nuclear volume 
fraction of about $1/8$ is smaller than the critical density for proton 
clustering.  We find that this critical density is in turn controlled by the 
symmetry energy at subnuclear densities.

     Earlier investigations on such exotic nuclei are more or less based on 
specific nuclear models \cite{PR,WS}.  An exception is the work by Watanabe et 
al.\ \cite{WIS} which is systematic in the sense that the liquid-drop model 
calculations were performed in a way dependent on the proton chemical 
potential in pure neutron matter, $\mu_p^{(0)}$, and the surface tension.  It 
was found that the density at which the system dissolves into uniform matter 
increases with increasing $\mu_p^{(0)}$.  However, it remains to be clarified 
why some nuclear models \cite{LRP,Chen,SLy} predict the absence of bubbles 
and nonspherical nuclei.  It is important to note that these models predict
relatively high pressure for pure neutron matter (or, equivalently, 
relatively small symmetry energy) at densities around half the normal nuclear 
density, while the work by Watanabe et al.\ \cite{WIS} uses a parametrization 
\cite{BBP} based on the microscopic calculations by Siemens and Pandharipande 
\cite{SP} as the EOS of pure neutron matter
and fix the density dependence of the symmetry energy.
This parametrization is consistent with the recent Green's function Monte 
Carlo (GFMC) calculations \cite{CMPR} at neutron densities up to about half 
the normal nuclear density.  
We will give a unified picture about the presence of bubbles and nonspherical 
nuclei by describing 
the pressure of pure neutron matter in terms of the density dependence 
of the symmetry energy.

    
     The size and shape of nuclei in the crust bear relevance to the 
thermal and rotational evolution of neutron stars.  This is because thermal 
conductivity and neutrino emissivity in the crust are controlled by 
electron-nucleus scattering \cite{YP}, while the motion of superfluid neutron 
vortices in the crust is affected by vortex-nucleus interactions \cite{EB92}.  
However, real crustal matter has to be accompanied by defects and impurities, 
which can play a more important role in the star's evolution \cite{Jones}.  In
a real neutron star, furthermore, the nuclear system is more or less out of 
equilibrium in the course of mass accretion onto the surface of the star
and the star's spin-down \cite{Sato,HZ,IS}.  Such disordered and 
nonequilibrium properties are beyond the scope of this paper.


     In Sec.\ \ref{sec:model}, we construct a model for inhomogeneous 
nuclear matter at subnuclear densities in a way dependent on the EOS of
nearly symmetric nuclear matter near the saturation point.  The equilibrium 
size and shape of nuclei at given density are then calculated from the model
constructed in Sec.\ \ref{sec:size}.  Section \ref{sec:clustering} is devoted
to evaluations of the critical density at which uniform matter becomes 
unstable against proton clustering.  Our conclusions are given in Sec.\ 
\ref{sec:conc}.

\section{Model for matter at subnuclear densities}
\label{sec:model}

    In this section, we construct a macroscopic model for zero-temperature, 
$\beta$-equilibrated, inhomogeneous nuclear matter at subnuclear densities.
This is an extension of Ref.\ \cite{OI} to the case in which a gas of dripped
neutrons is present, which is based on Ref.\ \cite{O}.
Here we focus on how macroscopic properties of the 
system depend on the EOS of nearly symmetric nuclear matter and, for 
simplicity, ignore various effects such as nucleon pairing effects \cite{FMH},
shell effects in inhomogeneous matter \cite{OY,MH}, fluctuation-induced 
displacements of nuclei and bubbles \cite{WIS}, and electron screening effects
\cite{WI}.

    The bulk energy per nucleon is an essential ingredient of the
macroscopic nuclear model.  We set this energy as 
\begin{eqnarray}
  w&=&\frac{3 \hbar^2 (3\pi^2)^{2/3}}{10m_n n}(n_n^{5/3}+n_p^{5/3})
   \nonumber \\ & & +(1-\alpha^2)v_s(n)/n+\alpha^2 v_n(n)/n,
\label{eos1}
\end{eqnarray}
where 
\begin{equation}
  v_s=a_1 n^2 +\frac{a_2 n^3}{1+a_3 n}
\label{vs}
\end{equation}
and
\begin{equation}
  v_n=b_1 n^2 +\frac{b_2 n^3}{1+b_3 n}
\label{vn}
\end{equation}
are the potential energy densities for symmetric nuclear matter and pure 
neutron matter, $n_n$ and $n_p$ are the neutron and proton number densities,
$n=n_n+n_p$, $\alpha=(n_n-n_p)/n$ is the neutron excess, and $m_n$ is the 
neutron mass.  Expressions (\ref{eos1})--(\ref{vn}) can well reproduce the 
microscopic calculations of symmetric nuclear matter and pure neutron matter 
by Friedman and Pandharipande \cite{FP} in the variational method.  In this 
method, the isospin dependence of asymmetric matter EOS is shown to be well 
approximated by Eq.\ (\ref{eos1}) \cite{LP}.  (Replacement of the proton mass 
$m_p$ by $m_n$ in the proton kinetic energy would make only a negligible 
difference.)  For the later purpose of roughly describing the nucleon 
distribution in a nucleus, we incorporate into the potential energy densities 
(\ref{vs}) and (\ref{vn}) a low density behavior $\propto n^2$ as expected 
from a contact two-nucleon interaction.  

   A set of expressions (\ref{eos1})--(\ref{vn}) is one of the simplest 
that reduces to the usual form (\ref{eos0}) in the limit of $n\to n_0$ and 
$\alpha\to0$,
\begin{equation}
    w=w_0+\frac{K_0}{18n_0^2}(n-n_0)^2+ \left[S_0+\frac{L}{3n_0}(n-n_0)
      \right]\alpha^2.
\label{eos0}
\end{equation}
Here $w_0$, $n_0$, and $K_0$ are the saturation energy, saturation density, and
incompressibility of symmetric nuclear matter.  The parameters $L$ and $S_0$ 
are associated with the density dependent symmetry energy coefficient $S(n)$: 
$S_0$ is the symmetry energy coefficient at $n=n_0$, and 
$L=3n_0(dS/dn)_{n=n_0}$ is the symmetry energy density derivative coefficient 
(hereafter referred to as the ``density symmetry coefficient'').  As the 
neutron excess increases from zero, the saturation point moves in the density 
versus energy plane (see, e.g., the dotted lines in Fig.\ 2).  This 
movement is determined mainly by the parameters $L$ and $S_0$.  Up to second 
order in $\alpha$, the saturation energy $w_s$ and density $n_s$ are given by 
\begin{equation}
  w_s=w_0+S_0 \alpha^2
\label{ws}
\end{equation}
and
\begin{equation}
  n_s=n_0-\frac{3 n_0 L}{K_0}\alpha^2.
\label{ns}
\end{equation}
The slope, $y$, of the saturation line near $\alpha=0$ is thus expressed as
\begin{equation}
 y=-\frac{K_0 S_0}{3 n_0 L}.
\label{slope}
\end{equation}

    We determine the parameters $a_1, \cdots, b_3$ in such a way that
the charge number, charge radius, and mass of stable nuclei calculated in a 
macroscopic nuclear model constructed in Ref.\ \cite{OI} are consistent with 
the empirical data.  In the course of this determination, we fix $b_3$, 
which controls the EOS of matter for large neutron excess and high density,
at 1.58632 fm$^3$.  This value was obtained by one of the authors \cite{O} in 
such a way as to reproduce the neutron matter energy of Friedman and 
Pandharipande \cite{FP}.  Change in this parameter would make no significant
difference in the determination of the other parameters and the final phase 
diagram.


    We describe macroscopic nuclear properties in a way dependent on the EOS
parameters $a_1, \cdots, b_3$ by using a Thomas-Fermi model \cite{OI}.  The 
essential point of this model is to write down the total energy of a nucleus 
of mass number $A$ and charge number $Z$
as a function of the density distributions $n_n({\bf r})$ and $n_p({\bf r})$ 
in the form
\begin{equation}
 E=E_b+E_g+E_C+Nm_n c^2+Zm_p c^2,
\label{e}
\end{equation}
where 
\begin{equation}
  E_b=\int d^3 r n({\bf r})w\left(n_n({\bf r}),n_p({\bf r})\right)
\label{eb}
\end{equation}
is the bulk energy,
\begin{equation}
  E_g=F_0 \int d^3 r |\nabla n({\bf r})|^2
\label{eg}
\end{equation}
is the gradient energy with adjustable constant $F_0$,
\begin{equation}
  E_C=\frac{e^2}{2}\int d^3 r \int  d^3 r' 
      \frac{n_p({\bf r})n_p({\bf r'})}{|{\bf r}-{\bf r'}|}
\label{ec}
\end{equation}
is the Coulomb energy, and $N=A-Z$ is the neutron number.  This functional 
allows us to connect the EOS and the nuclear binding energy through the bulk 
energy part $E_b$.  For simplicity we use the following parametrization for 
the nucleon distributions $n_i(r)$ $(i=n,p)$: 
\begin{equation}
  n_i(r)=\left\{ \begin{array}{lll}
  n_i^{\rm in}\left[1-\left(\displaystyle{\frac{r}{R_i}}\right)^{t_i}\right]^3,
         & \mbox{$r<R_i,$} \\
             \\
         0,
         & \mbox{$r\geq R_i.$}
 \end{array} \right.
\label{ni}
\end{equation}
This parametrization allows for the central density, half-density radius, and 
surface diffuseness for neutrons and protons separately.  In order to 
construct the nuclear model in such a way as to reproduce empirical masses and
radii of stable nuclei, we first extremized the binding energy with respect to
the particle distributions for fixed mass number, five EOS parameters, and 
gradient coefficient.  
Next, for various sets of the incompressibility and 
the density symmetry coefficient, we obtained the remaining three EOS 
parameters and the gradient coefficient by fitting the calculated optimal 
values of charge number, mass excess, root-mean-square (rms) charge radius to 
empirical data for stable nuclei on the smoothed $\beta$ stability line 
\cite{O}.  In the range of the parameters 
$0<L<160$ MeV and 180 MeV $<K_0<360$ MeV, as long as $y \lesssim -200$ MeV 
fm$^3$, we obtained a reasonable fitting to such data (see Fig.\ 1).
As a result of this fitting, the parameters $n_0$, $w_0$, $S_0$, and $F_0$
are constrained as $n_0=0.14$--0.17 fm$^{-3}$, $w_0=-16\pm1$ MeV, 
$S_0=25$--40 MeV, and $F_0=66\pm6$ MeV fm$^5$.
We remark that a negative $L$ is inconsistent with the fact that the size
of $A=17,20,31$ isobars deduced from the experimental values of the 
interaction cross section tends to increase with
neutron/proton excess \cite{isobar}.  This inconsistency can be seen from 
Eq.\ (\ref{ns}) which shows that the saturation density $n_s$ 
increases (and hence the isobar size decreases) 
with neutron/proton excess for a negative $L$.

\begin{figure}[t]
\includegraphics[width=8.5cm]{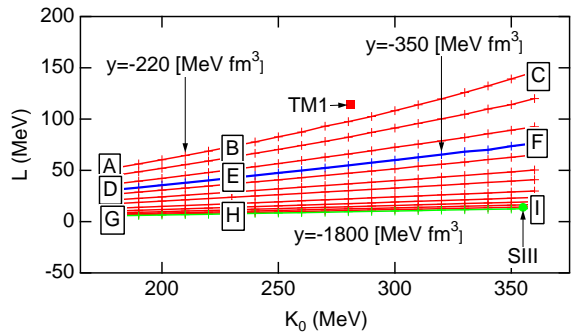}
\vspace{-0.5cm}
\caption{\label{lk0} (Color online)
The sets of $(L,K_0)$ (crosses) consistent with the mass 
and radius data for stable nuclei.  The thin lines are lines of constant $y$.
The labels A--I denote the sets for which we perform detailed calculations of 
the ground state properties of inhomogeneous nuclear matter at subnuclear 
densities.  For comparison, the values calculated from two mean-field models 
[TM1 (square) and SIII (dot)], which are known to be extreme cases \cite{OT}, 
are plotted.  The plot shows that our sets of $(L,K_0)$
effectively cover such extreme cases.
}
\end{figure}

     We remark that in this range the calculations agree well with a more 
extended data set of nuclear masses for $A\geq2$ \cite{Audi} and charge radii 
for $A\geq50$ \cite{dV}.  The rms deviations of the calculated masses from the 
measured values are $\sim3$--5 MeV, which are comparable with the deviations 
obtained from a Weizs{\" a}cker Bethe formula, while the rms deviations of the 
calculated charge radii from the measured values are about 0.06 fm, which 
are comparable with the deviations obtained from the $A^{1/3}$ law.

     Let us summarize the macroscopic nuclear model used here.  This model 
can describe global nuclear properties such as masses and rms radii in a manner
that is dependent on the EOS of nuclear matter.  One of the important
predictions of this model was that the matter radii depend appreciably on the 
density symmetry coefficient $L$, while being almost independent of the 
incompressibility $K_0$.  
Although the present macroscopic approach has some limitations
in describing the nuclear surface, it is still useful for examining
the phase diagram of nuclear matter at subnuclear densities \cite{O}.


     For the purpose of describing matter in neutron star crusts, we proceed 
to extend the above-described nuclear model to the case of nuclei of 
various shapes embedded in a gas of dripped neutrons by following a line of
arguments of Ref.\ \cite{O}.  Here we also take into account a gas of 
electrons as a constituent of matter in the crust and impose $\beta$ stability
and charge neutrality in the system.  

     We consider five phases that consist of spherical nuclei, cylindrical 
nuclei, planar nuclei, cylindrical bubbles, and spherical bubbles, 
respectively.  Each phase is assumed to be composed of a Coulomb lattice of a 
single species of nucleus or bubble at a given baryon density $n_b$.  
For the convenience of practical calculations, we adopt the 
Wigner-Seitz approximation.  In this approximation, a cell in the bcc lattice, 
including a spherical nucleus or bubble, is replaced by a Wigner-Seitz cell 
defined as a sphere having the same volume ($a^3$) and center.  We refer to 
$a$ as the lattice constant.  A cylindrical nucleus
or bubble having an infinitely long axis and a circular section is taken to be
contained in a cylindrical Wigner-Seitz cell having the same axis
in place of a cell in the two-dimensional triangular lattice.  For a planar 
nucleus, a Wigner-Seitz cell is identical with a cell in the one-dimensional 
layered lattice.  For the sake of convenience, we redefine the cylindrical 
and slab Wigner-Seitz cells as a cylinder of height $a$ and base area $a^2$ 
and a slab of thickness $a$ and surface area $a^2$, respectively
(see Figs. 1 and 2 in Ref.\ \cite{O}).

    For each unit cell, we write the total energy as
\begin{equation}
  W=W_N+W_e+W_C,
   \label{w}
\end{equation}
where $W_N$, $W_e$, and $W_C$ are the nuclear energy, the electron energy,
and the Coulomb energy.  

     As in Eq.\ (\ref{e}), the nuclear energy is again 
expressed in the density functional form:
\begin{eqnarray}
  W_N&=&\int_{\rm cell} d^3 r 
    [n({\bf r})w\left(n_n({\bf r}),n_p({\bf r})\right)
   \nonumber \\ &&
     +m_n c^2 n_n({\bf r})+m_p c^2 n_p({\bf r})
     +F_0 |\nabla n({\bf r})|^2].
   \label{wn}
\end{eqnarray}
For a spherical nucleus in vacuum, this expression reduces to $E-E_C$
[see Eq.\ (\ref{e})].

     The electron energy can be approximated as the energy of an ideal
uniform Fermi gas,
\begin{equation}
    \frac{W_e}{a^3}=\frac{m_e^4c^5}{8\pi^2\hbar^3}
        \{x_e(2x_e^2+1)(x_e^2+1)^{1/2}-\ln[x_e+(x_e^2+1)^{1/2}]\}
    \label{we}
\end{equation}
with
\begin{equation}
    x_e=\frac{\hbar (3\pi^2 n_e)^{1/3}}{m_e c},
    \label{xe}
\end{equation}
where $m_e$ is the electron mass, and $n_e$ is the electron number density
that satisfies the charge neutrality condition,
\begin{equation}
    a^3 n_e= \int_{\rm cell} d^3 r n_p({\bf r}).
    \label{cn}
\end{equation}
We remark that $n_e$ is so high that we can safely ignore inhomogeneity of 
the electron density induced by the electron screening of nuclei or bubbles 
\cite{WI} and the Hartree-Fock corrections to the electron energy.

    The Coulomb energy is composed of the proton self-Coulomb energy and 
the lattice energy.  We write the Coulomb energy as
\begin{equation}
  W_C=\frac12 \int_{\rm cell} d^3 r e[n_p({\bf r})-n_e]\phi({\bf r})
      +\Delta W_1,
    \label{wc}
\end{equation}
where $\phi({\bf r})$ is the electrostatic potential in a Wigner-Seitz
cell, and $\Delta W_1$ is the difference of the rigorous calculation 
\cite{OHY} for a cell in the bcc (triangular) lattice of spherical 
(cylindrical) nuclei or bubbles having sharp surfaces from the 
Wigner-Seitz value, as parametrized in Ref.\ \cite{O}.  We take 
into account $\Delta W_1$, which is a less than 1 \% correction,
because $\Delta W_1$ depends sensitively on the dimensionality of the 
lattice. (Note that $\Delta W_1=0$ for the layered lattice of slab nuclei.)

     For nucleon distributions in the Wigner-Seitz cell, we simply 
generalize the parametrization (\ref{ni}) for a nucleus in vacuum into
\begin{equation}
  n_i(r)=\left\{ \begin{array}{lll}
  (n_i^{\rm in}-n_i^{\rm out})
  \left[1-\left(\displaystyle{\frac{r}{R_i}}\right)^{t_i}\right]^3
          +n_i^{\rm out},
         & \mbox{$r<R_i,$} \\
             \\
         n_i^{\rm out},
         & \mbox{$R_i\leq r.$}
 \end{array} \right.
\label{nig} 
\end{equation}
Here $r$ is the distance from the central point, axis, or plane of
the unit cell.  In the case of nuclei, $n_p^{\rm out}=0$, while
in the case of bubbles, $n_p^{\rm in}=0$.

     We finally determine the equilibrium configuration of the system
at given baryon density,
\begin{equation}
  n_b=a^{-3}\int_{\rm cell} d^3 r n({\bf r}).
 \label{nb}
\end{equation}
First, for each of the five inhomogeneous phases, we minimize 
the total energy density $W/a^3$ with respect to the eight
parameters $a$, $n_n^{\rm in}$, $n_n^{\rm out}$, $n_p^{\rm in}$ (for nuclei)
or $n_p^{\rm out}$ (for bubbles), $R_n$, $R_p$, $t_n$, and $t_p$.
This minimization implicitly allows for the stability of the nuclear matter 
region (the region containing protons) with respect to change in the size, 
neutron drip, $\beta$-decay, and pressurization.  In addition to
the five inhomogeneous
phases, we consider a uniform phase of $\beta$-equilibrated, neutral nuclear 
matter.  The energy density of this phase is the sum of the nucleon part 
$nw+m_n c^2 n_n+m_p c^2 n_p$ [see Eq.\ (\ref{eos1})] and the electron part 
(\ref{we}).  By comparing the resultant six energy densities, we can 
determine the equilibrium phase.


\section{Equilibrium size and shape of nuclei}
\label{sec:size}

     We proceed to show the results for the equilibrium nuclear matter
configuration obtained for various sets of the EOS parameters $L$ and $K_0$
as shown in Fig.\ 1.  These parameters are still uncertain since they are 
little constrained from the mass and radius data for stable nuclei \cite{OI}.  
As we shall see, the charge number of spherical nuclei and the density region 
containing bubbles and nonspherical nuclei have a strong correlation with $L$.

     We first focus on spherical nuclei, which constitute an equilibrium
state in the low density region.  We calculate the charge number of 
the equilibrium nuclide as a function of $n_b$ for the EOS models A--I as 
depicted in Fig.\ 2.  Note that the recent GFMC calculations of the energy of 
neutron matter based on the Argonne v8' potential \cite{CMPR} are close to the 
behavior of the model E.   Hereafter we will thus call the model E as a 
typical one.  The result is shown in Fig.\ 3.  For densities below
$\sim0.01$ fm$^{-3}$, the calculated density dependence of the charge number 
$Z$ is almost flat, a feature consistent with the results in earlier 
investigations \cite{PR}.  More important, the calculated charge number is 
larger for the EOS models having smaller $L$, and this difference in $Z$ is 
more remarkable at higher densities.

\begin{figure}[t]
\includegraphics[width=8.5cm]{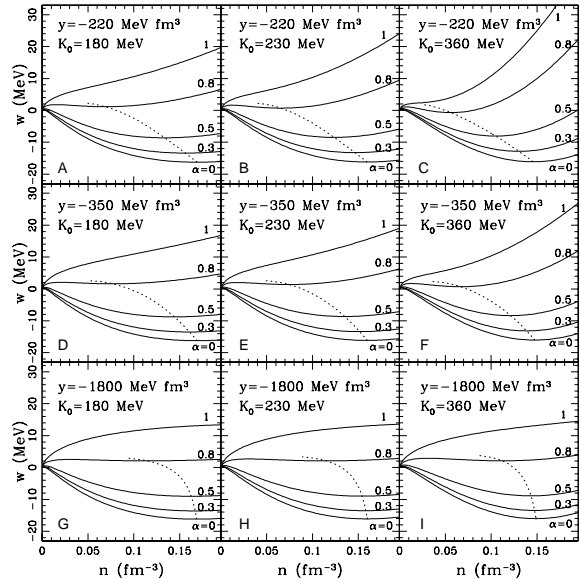}
\vspace{-0.5cm}
\caption{\label{eos}Energy per nucleon of nuclear matter for the nine sets 
of $(L,K_0)$ referred to as A--I in Fig.\ 1.  In each panel, the solid lines
are the energy at $\alpha=0, 0.3, 0.5, 0.8, 1$, and the dotted line is the 
saturation line.}
\end{figure}

\begin{figure}[t]
\includegraphics[width=8.5cm]{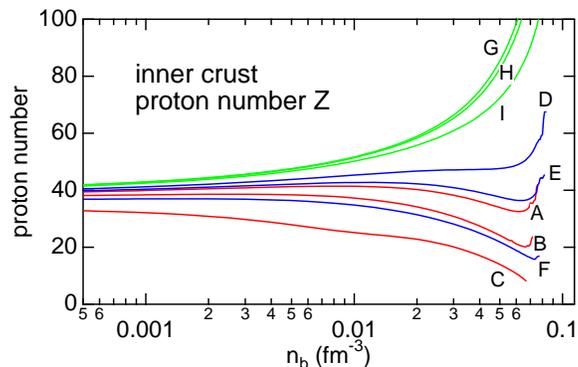}
\vspace{-0.5cm}
\caption{\label{z} (Color online)
The charge number of spherical nuclei as a function of
$n_b$, calculated for the EOS models A--I.}
\end{figure}

     As we shall see later in Fig.\ 6, this property of $Z$ is related 
to the tendency that with increasing $L$, the nuclear density decreases 
while the density of 
the neutron gas increases.   Note that $Z$ is, within a liquid-drop model 
\cite{PR}, determined by the size equilibrium condition relating the Coulomb 
and surface energies in such a way that $Z$ increases with increasing surface 
tension.  Since the Thomas-Fermi model adopted here can be mapped onto a 
compressible liquid-drop model \cite{OI}, the present results may well be 
interpreted in terms of the liquid-drop model.  In fact we shall estimate 
the surface tension from the Thomas-Fermi model as a function of $L$
and discuss how the surface tension depends on the nuclear density and the 
neutron sea density.

     We also note that the density at which the phase with spherical 
nuclei ceases to be in the ground state is between 0.05 fm$^{-3}$ and 
0.07 fm$^{-3}$.  This result, consistent with the results obtained in earlier 
investigations \cite{PR,O,WIS}, will be discussed below in terms of 
fission instability.

     The average proton fraction, which is the charge number divided by
the total nucleon number in the cell,
is plotted in Fig.\ 4.  We observe that the dependence of
the average proton fraction on the EOS models is similar to that of $Z$.
We also find that the average proton fraction basically decreases with baryon 
density.  This is a feature coming from the fact that as the baryon density 
increases, the electron chemical potential increases under charge neutrality
and then the nuclei become more neutron-rich under weak equilibrium.

\begin{figure}[t]
\includegraphics[width=8.5cm]{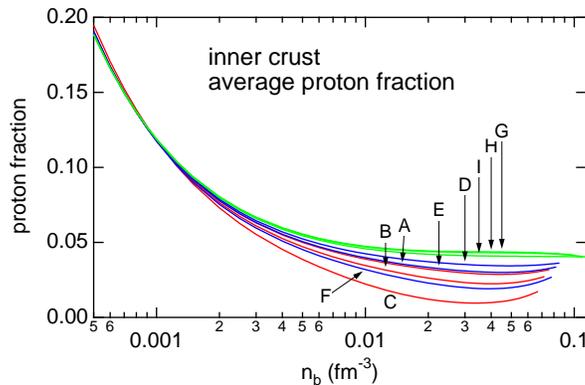}
\vspace{-0.5cm}
\caption{\label{x} (Color online)
The average proton fraction as a function of
$n_b$, calculated for the EOS models A--I.}
\end{figure}

     We next consider the density region where bubbles and nonspherical nuclei
appear in equilibrium, i.e., the density region of the ``pasta'' phases.
We start with such a density region calculated for the
EOS models A--I.  The results are plotted in Fig.\ 5.  Except for the model C,
we obtain the successive first order transitions with increasing density:
sphere $\to$ cylinder $\to$ slab $\to$ cylindrical hole $\to$ spherical hole
$\to$ uniform matter.  A marked correlation of 
the upper end of the density region with the parameter $L$ can be observed 
by referring to Fig.\ 1.  This dependence will be examined in detail in the 
next section.

\begin{figure}[t]
\includegraphics[width=8.5cm]{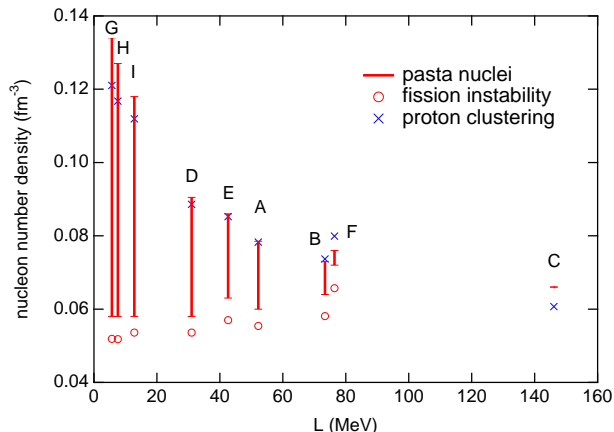}
\vspace{-0.5cm}
\caption{\label{pasta} (Color online)
The density region containing bubbles and nonspherical
nuclei as a function of $L$, calculated for the EOS models A--I.  For
comparison, the density corresponding to $u=1/8$ in the phase with spherical 
nuclei and the onset density, $n(Q)$, of proton clustering in uniform 
nuclear matter, which will be discussed in Sec.\ \ref{sec:clustering}, 
are also plotted by circles and crosses, respectively.}
\end{figure}

     As can be seen in Fig.\ 5, the lower end of the density region of the
pasta phases has only a weak dependence on the EOS models.  In order to have
a closer look at this feature, it is instructive to calculate a density at 
which spherical nuclei become susceptible to fission-inducing quadrupolar 
deformations.  Within the framework of a liquid-drop model, a spherical 
liquid-drop in a Wigner-Seitz cell is predicted to
undergo such a fission-like instability 
when the volume fraction $u$, i.e., the ratio of the liquid-drop volume to the
cell volume, becomes approximately $1/8$ \cite{PR}.  In such a closely packed
situation, the Coulomb self-energy of the liquid-drop amounts to twice the 
surface energy even under size equilibrium.  We note that the density 
corresponding to $u=1/8$ is generally within $\pm0.01$ fm$^{-3}$ of the 
transition density calculated from the energy comparison between the phases 
with spherical nuclei and with cylindrical nuclei.


     In the present model, we evaluate the volume fraction $u$ as 
$4\pi(r_p/a)^3 /3$, where $r_p$ is the rms radius of the 
proton distribution multiplied by a factor $\sqrt{5/3}$.  This is because 
the proton self-Coulomb energy is relevant to the 
fission-like instability.  At $u=1/8$, we calculate the equilibrium 
properties of matter with spherical nuclei for the parameter sets $(L,K_0)$ 
included in Fig.\ 1.  The results are plotted as a function of $L$ 
in Fig.\ 6.  The results for $L>100$ MeV are scarce since in this case the 
pressure of neutron matter is too high for $u$ to amount to $1/8$.  We 
remark that the results show only a weak dependence on $K_0$. 

\begin{figure}[t]
 \begin{minipage}{.45\linewidth}
  \includegraphics[width=8.5cm]{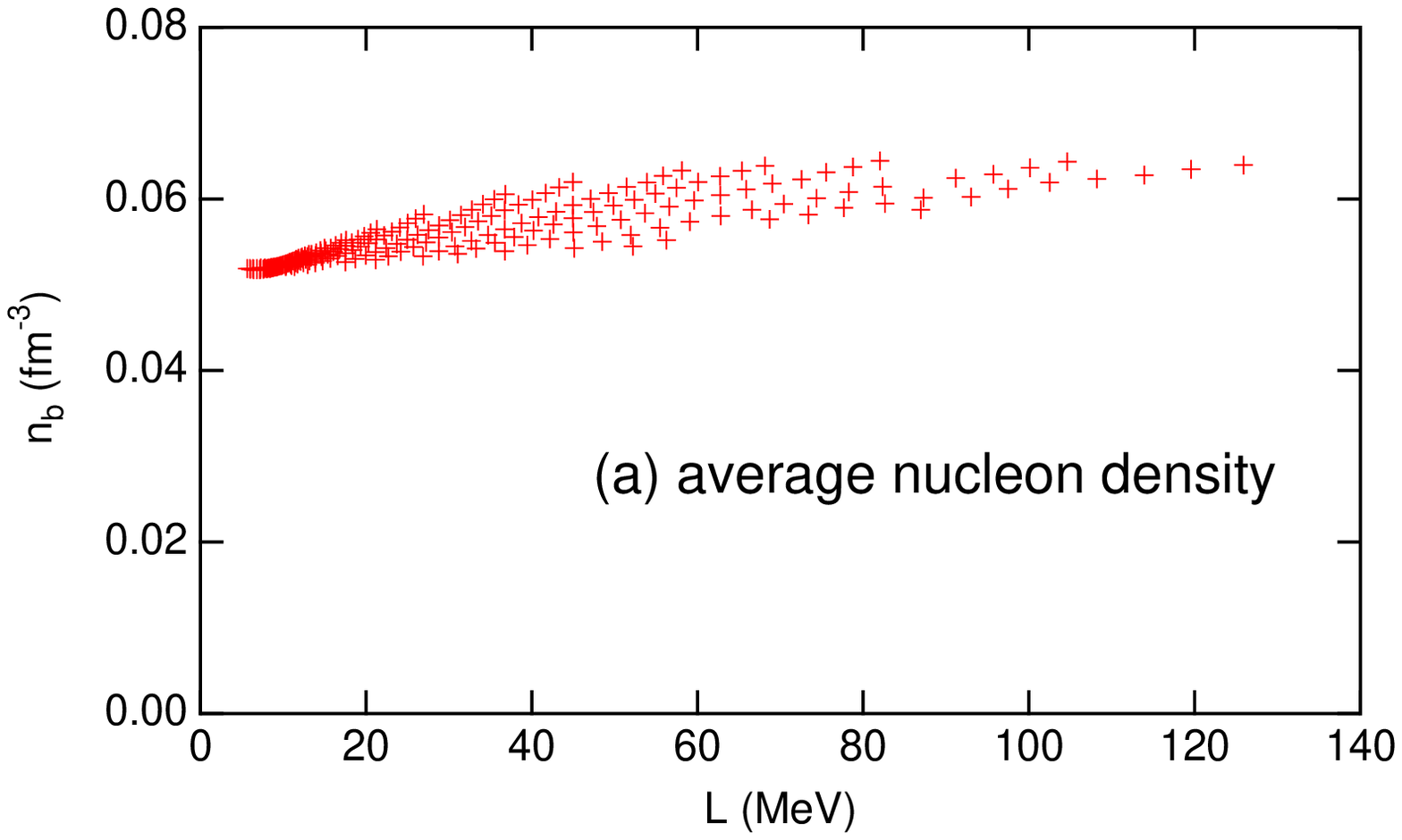}  
 \end{minipage}
 \begin{minipage}{.45\linewidth}
  \includegraphics[width=8.5cm]{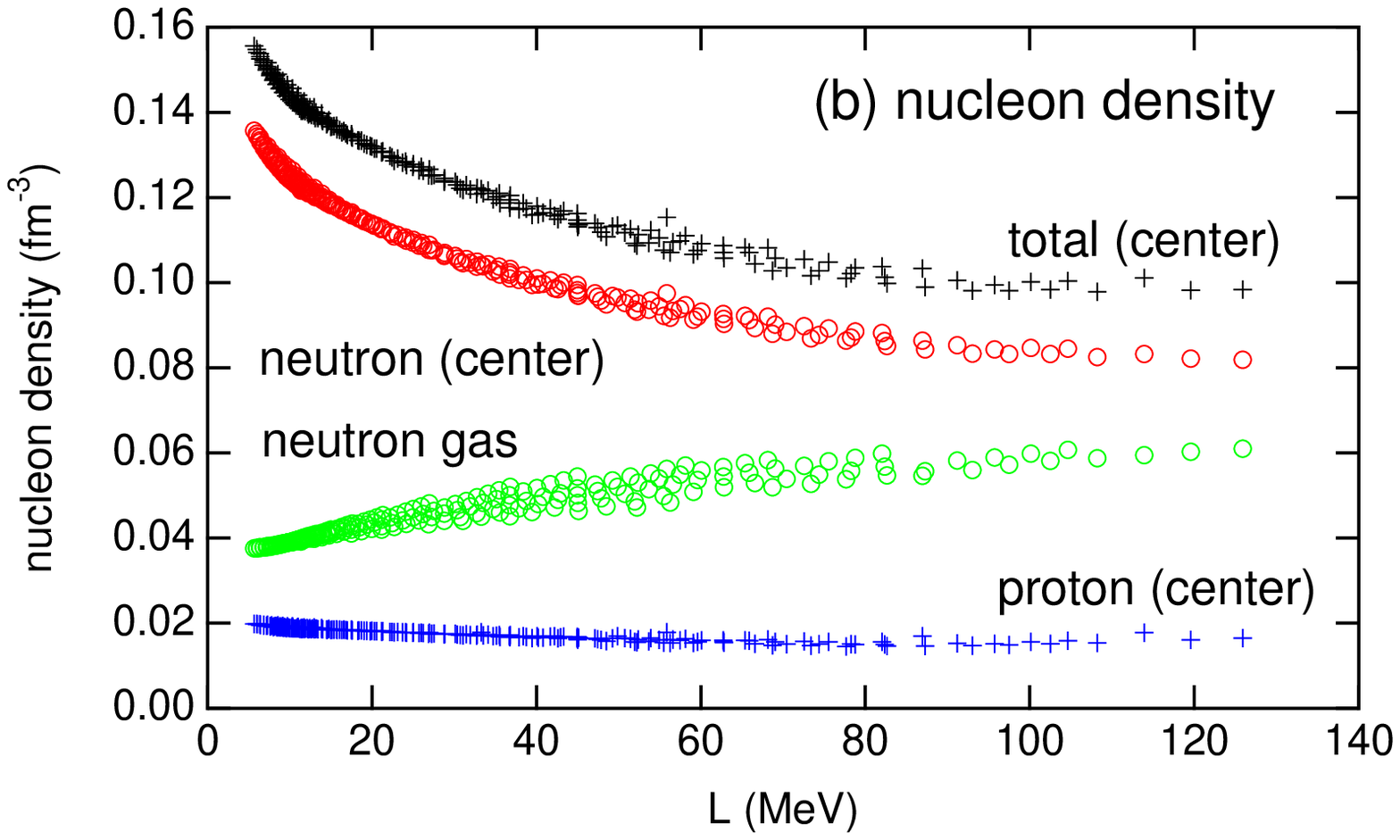}
 \end{minipage}
 \begin{minipage}{.45\linewidth}
  \includegraphics[width=8.5cm]{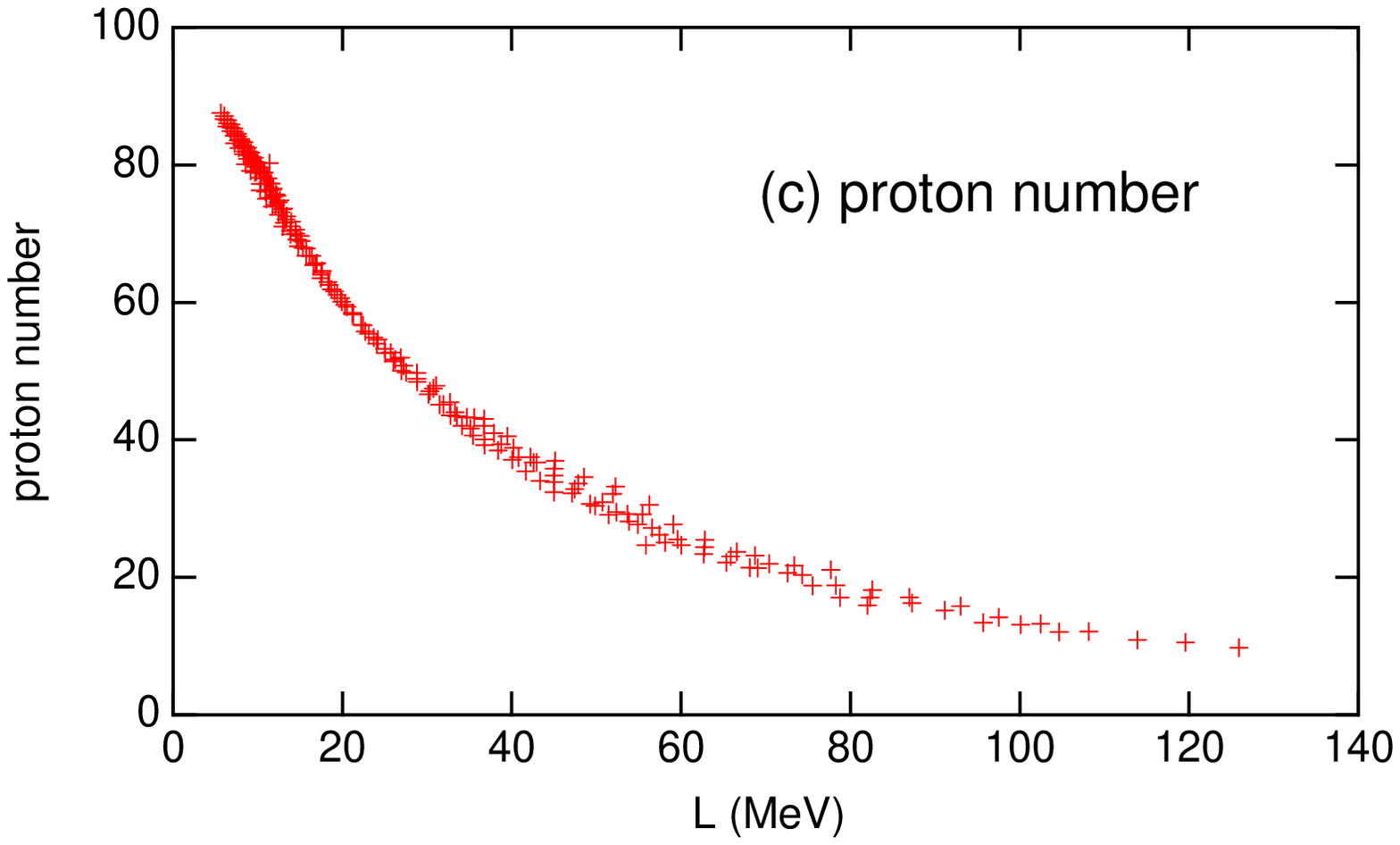}
 \end{minipage}
 \begin{minipage}{.45\linewidth}
  \includegraphics[width=8.5cm]{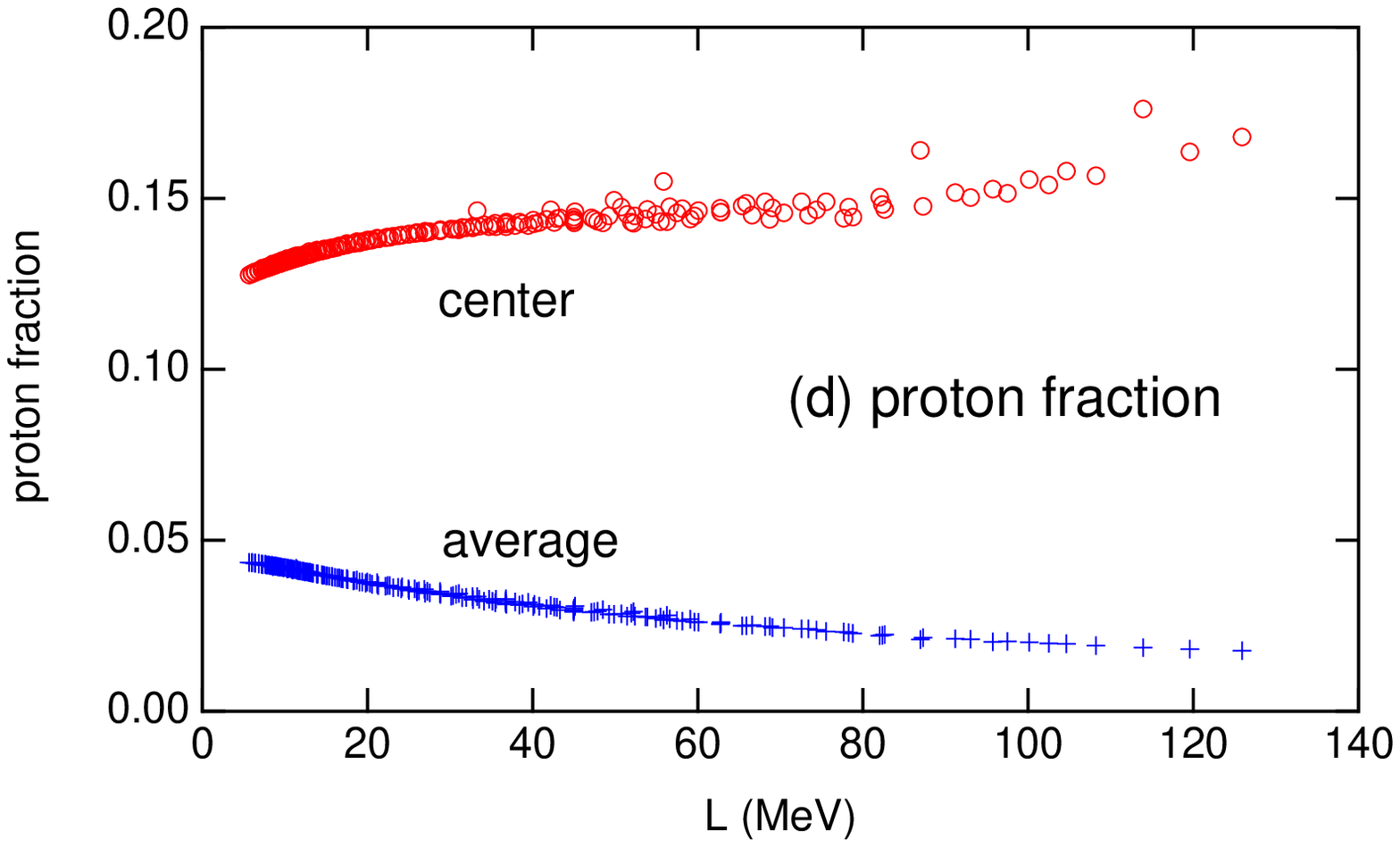}
 \end{minipage}
 \begin{minipage}{.45\linewidth}
  \includegraphics[width=8.5cm]{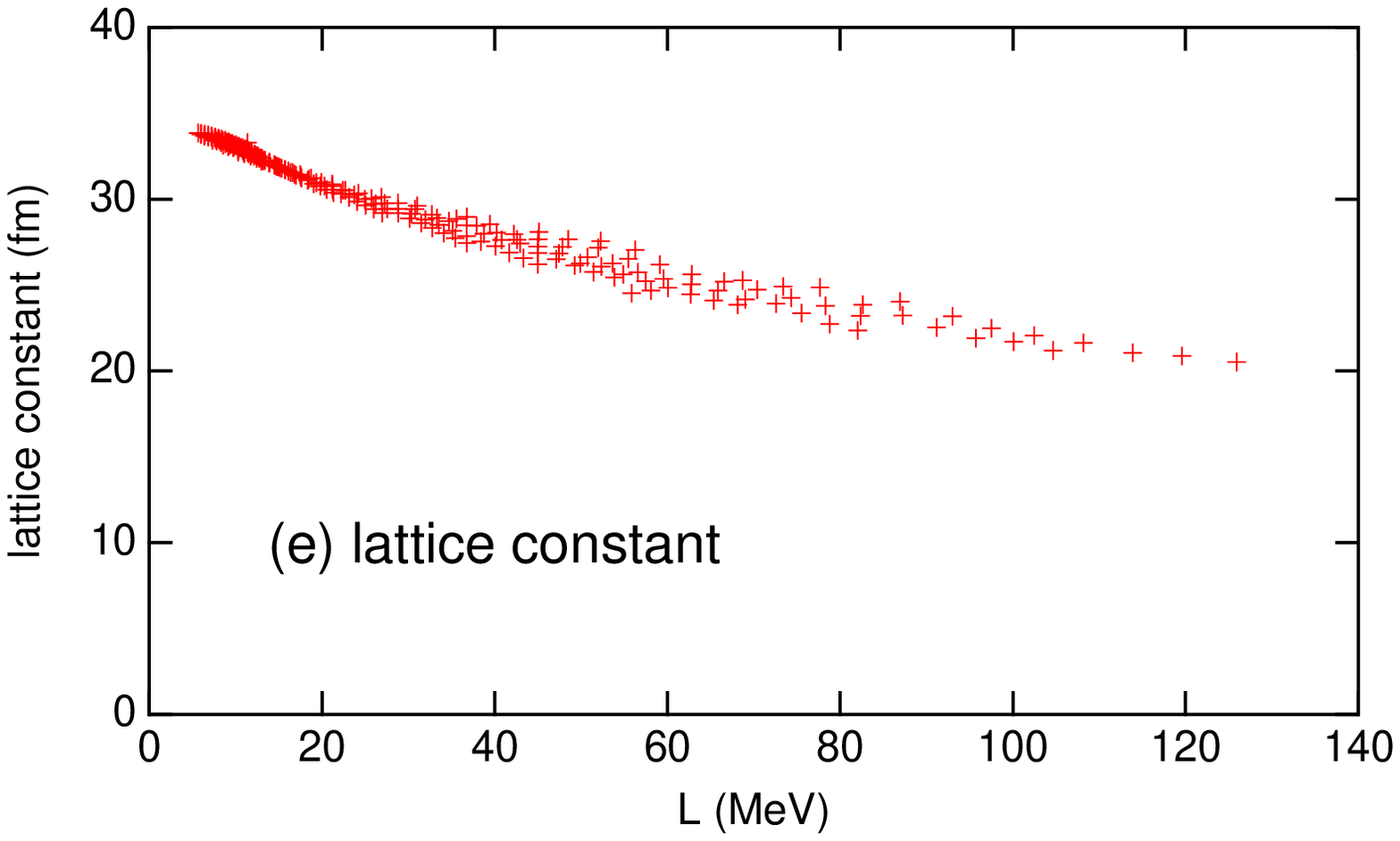}
 \end{minipage}
 \begin{minipage}{.45\linewidth}
  \includegraphics[width=8.5cm]{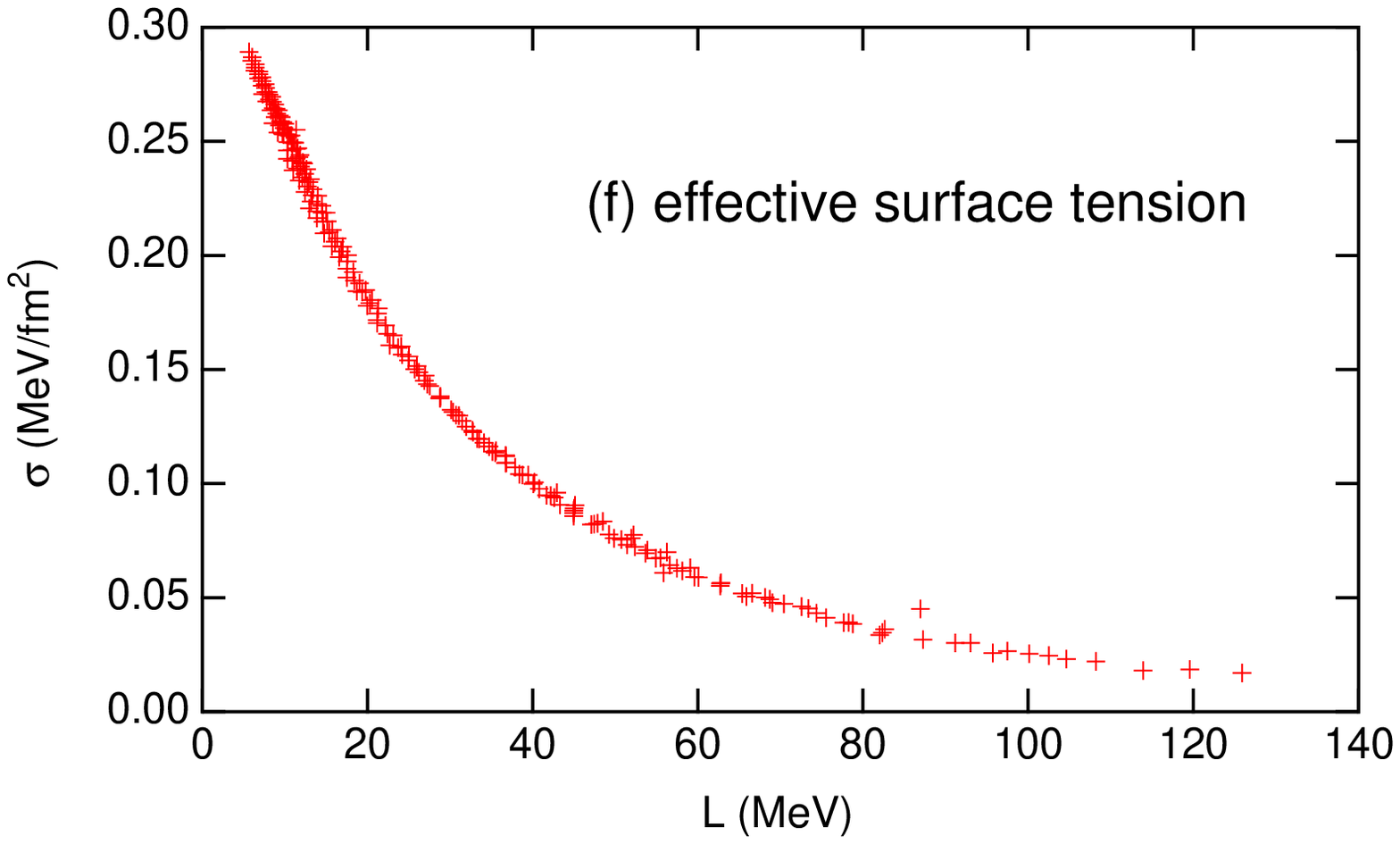}
 \end{minipage}
 \caption{\label{1/8} (Color online)
The equilibrium properties of matter with spherical 
nuclei as a function of $L$, calculated at fixed volume fraction $u=1/8$.
The baryon density $n_b$ (a), the nucleon densities in the center and boundary 
of the cell (b), the nuclear charge number $Z$ (c), 
the central and average proton fractions (d), 
the lattice constant $a$ (e), and 
the effective surface tension $\sigma$ (f) 
are plotted.
}
\end{figure}

     It is important to note that the baryon density at $u=1/8$ is almost flat
at $\sim0.06$ fm$^{-3}$ [see Fig.\ 6(a)].  This is consistent with the lower
end of the density region of the pasta phases as depicted in Fig.\ 5.  This 
magnitude of $n_b$ at $u=1/8$ can be roughly understood from a simple formula 
$n_b\simeq(n_p^{\rm in}+n_n^{\rm in})u+n_n^{\rm out}(1-u)$ with the values of 
$n_p^{\rm in}$, $n_n^{\rm out}$, and $n_n^{\rm out}$ in Fig.\ 6(b).  
We also note that with increasing $L$, the central density decreases
while the neutron sea density increases.  This is natural because 
both the saturation density of nuclear matter at nonzero neutron excess 
and the symmetry energy at subnuclear densities decreases with $L$.

     We now turn to the $L$ dependence of the charge number at
$u=1/8$ [see Fig.\ 6(c)].  The charge number decreases with $L$. 
This feature can be understood from the size equilibrium 
condition within a liquid-drop picture \cite{PR}.  This condition states that 
the Coulomb energy of a cell is half as large as the nuclear surface energy.
Consequently, the equilibrium charge number squared is proportional to the 
surface tension and to the nuclear volume.  Note that the proton fraction
in the nuclear center and the nuclear volume have 
a relatively weak
dependence on $L$ [see Figs.\ 6(d) and 6(e)].  The surface tension is thus 
expected to have 
a similar $L$ dependence to that of the charge number squared, through
the dependence on the densities inside and outside the nucleus.  From 
the Thomas-Fermi model, it is reasonable to estimate the effective
surface tension as 
\begin{equation}
  \sigma=\frac{W_g}{2\pi r_p^2},
\end{equation}
where 
\begin{equation}
  W_g=\int_{\rm cell} d^3 r F_0 |\nabla n({\bf r})|^2,
\end{equation}
is the gradient energy per cell.  This is because in equilibrium, the 
Coulomb energy of a cell is as large as $W_g$ \cite{O}, implying that the 
nuclear surface energy, $\approx 4\pi\sigma r_p^2$, is twice as large 
as $W_g$.  The surface tension thus estimated 
basically follows the behavior of 
$Z^2$, as can be seen from Figs.\ 6(c) and 6(f).  The surface tension is 
generally the function of the neutron excess in the nuclear interior and the 
densities inside and outside the nucleus \cite{BBP}.  Since the density 
gradient in the surface region tends to become small for smaller difference 
between the central density and the neutron sea density, the surface tension
decreases with $L$ as shown in Fig.\ 6(f).



\section{Proton clustering in uniform matter}
\label{sec:clustering}

     In this section, we focus on the upper end of the density region of the
pasta phases.  This upper end corresponds roughly to a density at which 
uniform nuclear matter neutralized and $\beta$ equilibrated by electrons
becomes unstable against proton clustering.  In fact, this correspondence 
can be seen from Fig.\ 5.

     We calculate the onset density of proton clustering by following a 
line of argument of Baym, Bethe, and Pethick \cite{BBP}.  This density was 
obtained in Ref.\ \cite{BBP} by expanding the energy density functional
$E[n_{i}({\bf r})]$ ($i=n,p,e$) of the system with respect to small 
density fluctuations $\delta n_i({\bf r})$ around the homogeneous state.
While the contribution of first order in $\delta n_i({\bf r})$ vanishes due 
to equilibrium of the unperturbed homogeneous system, the second order
contribution can be described in the spirit of the Thomas-Fermi model
used here as
\begin{equation}
E-E_0=\frac12 \int \frac{d^3 q}{(2\pi)^3}v(q)|\delta n_p({\bf q})|^2,
\end{equation}
where $E_0$ is the ground-state energy, $\delta n_p({\bf q})$ is the Fourier
transform of $\delta n_p({\bf r})$, and $v(q)$ is the potential of the 
effective interaction between protons as given by 
\begin{equation}
v(q)=v_0+\beta q^2+\frac{4\pi e^2}{q^2 + k_{\rm TF}^2}.
\end{equation}
Here, 
\begin{equation}
v_0=\frac{\partial\mu_p}{\partial n_p}
   -\frac{(\partial\mu_p/\partial n_n)^2}{\partial\mu_n/\partial n_n},
   \label{v0}
\end{equation}
\begin{equation}
\beta=2F_0(1+2\zeta+\zeta^2),
\end{equation}
\begin{equation}
\zeta=-\frac{\partial\mu_p/\partial n_n}{\partial\mu_n/\partial n_n},
\end{equation}
with $\mu_{n (p)}$ the neutron (proton) chemical potential, and
$k_{\rm TF}\approx0.3n_e^{1/3}$ is the inverse of the Thomas-Fermi screening 
length of the electron gas.  The effective potential $v(q)$ takes a minimum 
value $v_{\rm min}$ at $q=Q$, where 
\begin{equation}
Q^2=\left(\frac{4\pi e^2}{\beta}\right)^{1/2}-k_{\rm TF}^2,
\end{equation}
\begin{equation}
v_{\rm min}=v_0+2(4\pi e^2 \beta)^{1/2} -\beta k_{\rm TF}^2.
\end{equation}

     In the energy expansion up to second order in $\delta n_i$, the condition
that uniform nuclear matter becomes unstable with respect to proton clustering
reads $v_{\rm min}=0$.  Generally, $v_{\rm min}$ is dominated by the bulk 
contribution $v_0$, which decreases with decreasing density (see Fig.\ 7).
This density dependence ensures the presence of a critical density, $n(Q)$, 
above (below) which the matter is stable (unstable) with respect to proton 
clustering.  Hereafter we will estimate $n(Q)$ without including the gradient 
and Coulomb contributions to $v_{\rm min}$, which act to reduce $n(Q)$ only 
by an amount of order 0.02 fm$^{-3}$ (see Fig.\ 7 and also Ref.\ \cite{IWS}).

     The density dependence of $v_0$ can be seen by substituting Eq.\ 
(\ref{eos1}) into Eq.\ (\ref{v0}).  In the limit of $\alpha\to1$, to which
nuclear matter $\beta$ equilibrated and neutralized by the electron 
gas is close at subnuclear densities (see Fig.\ 7), $v_0$ behaves roughly as
\begin{equation}
  v_0 \sim 8\frac{(\partial\mu/\partial n)_{\alpha=0}}
                {(\partial\mu/\partial n)_{\alpha=1}}
           \frac{S(n)}{n},
\end{equation}
with the symmetry energy coefficient $S(n)=w_{\alpha=1}(n)-w_{\alpha=0}(n)$.
At subnuclear densities, $S(n)/n$ depends only weakly on $n$, while the 
compressibility ratio between pure neutron matter and symmetric nuclear 
matter, $(\partial\mu/\partial n)_{\alpha=0}
/(\partial\mu/\partial n)_{\alpha=1}$, increases almost linearly with density
because of the saturation property of the symmetric nuclear matter.

\begin{figure}[t]
\begin{minipage}{.45\linewidth}
 \includegraphics[width=8.5cm]{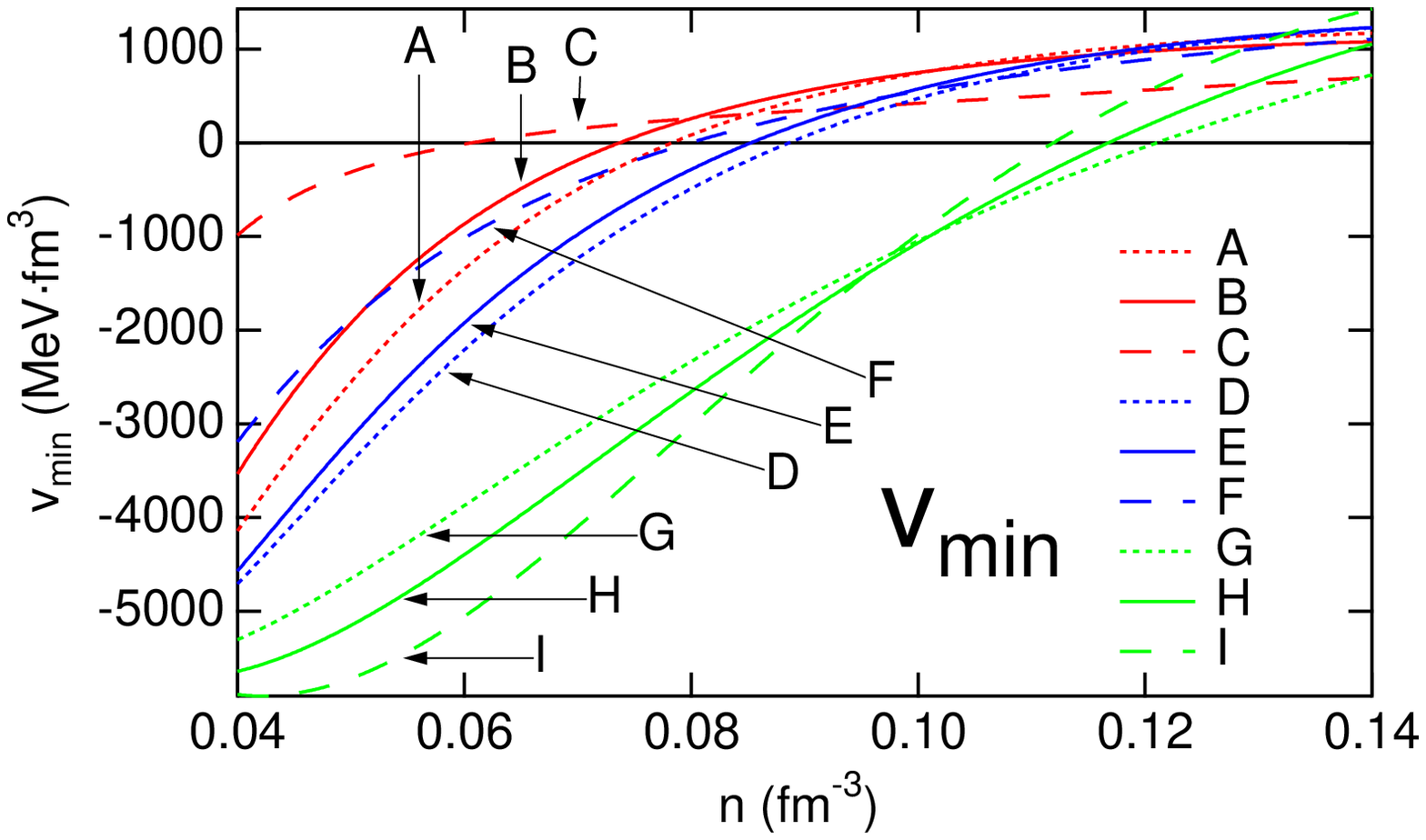}  
\end{minipage}
\begin{minipage}{.45\linewidth}
 \includegraphics[width=8.5cm]{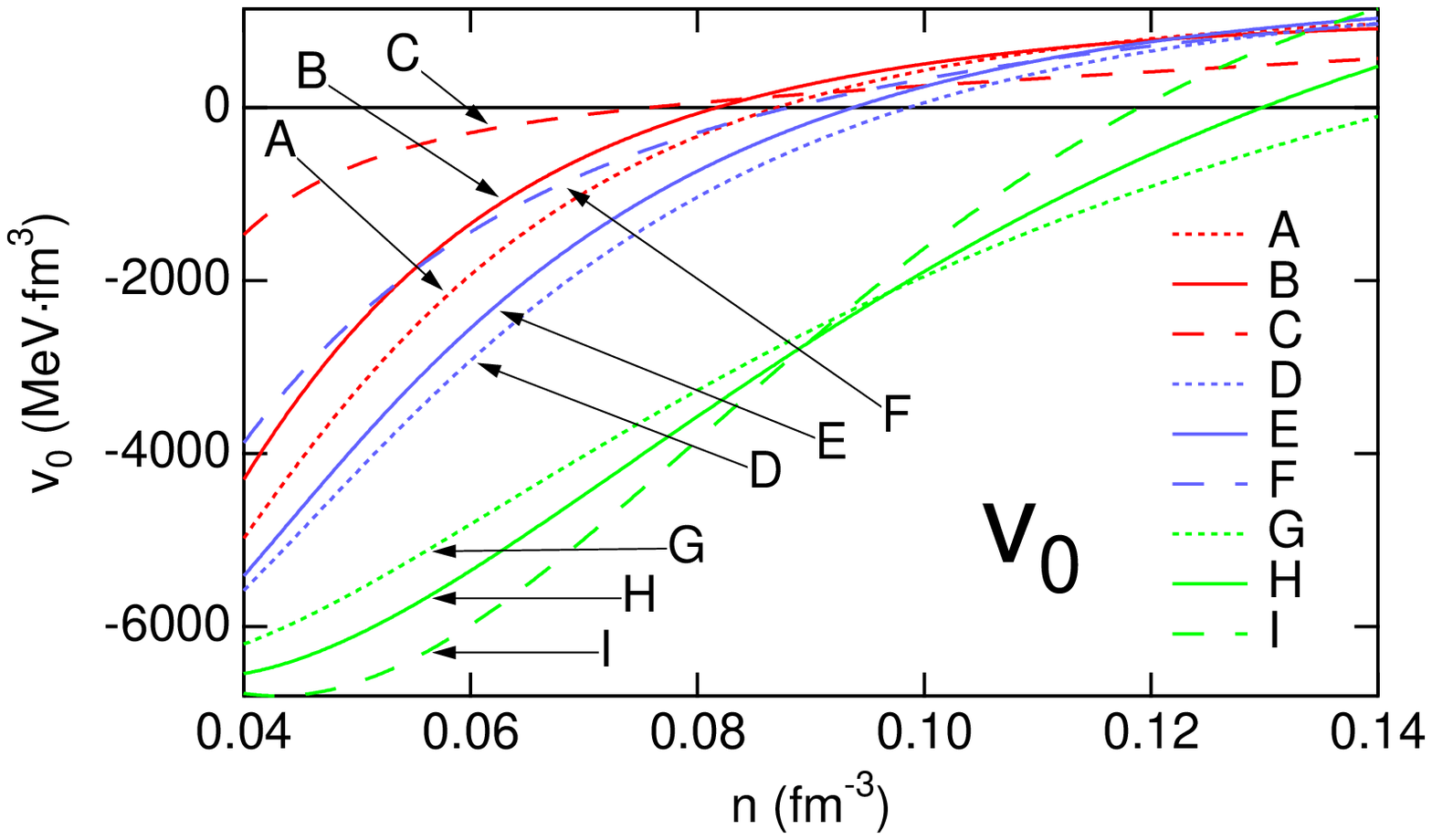}
\end{minipage}
\begin{minipage}{.45\linewidth}
 \includegraphics[width=8.5cm]{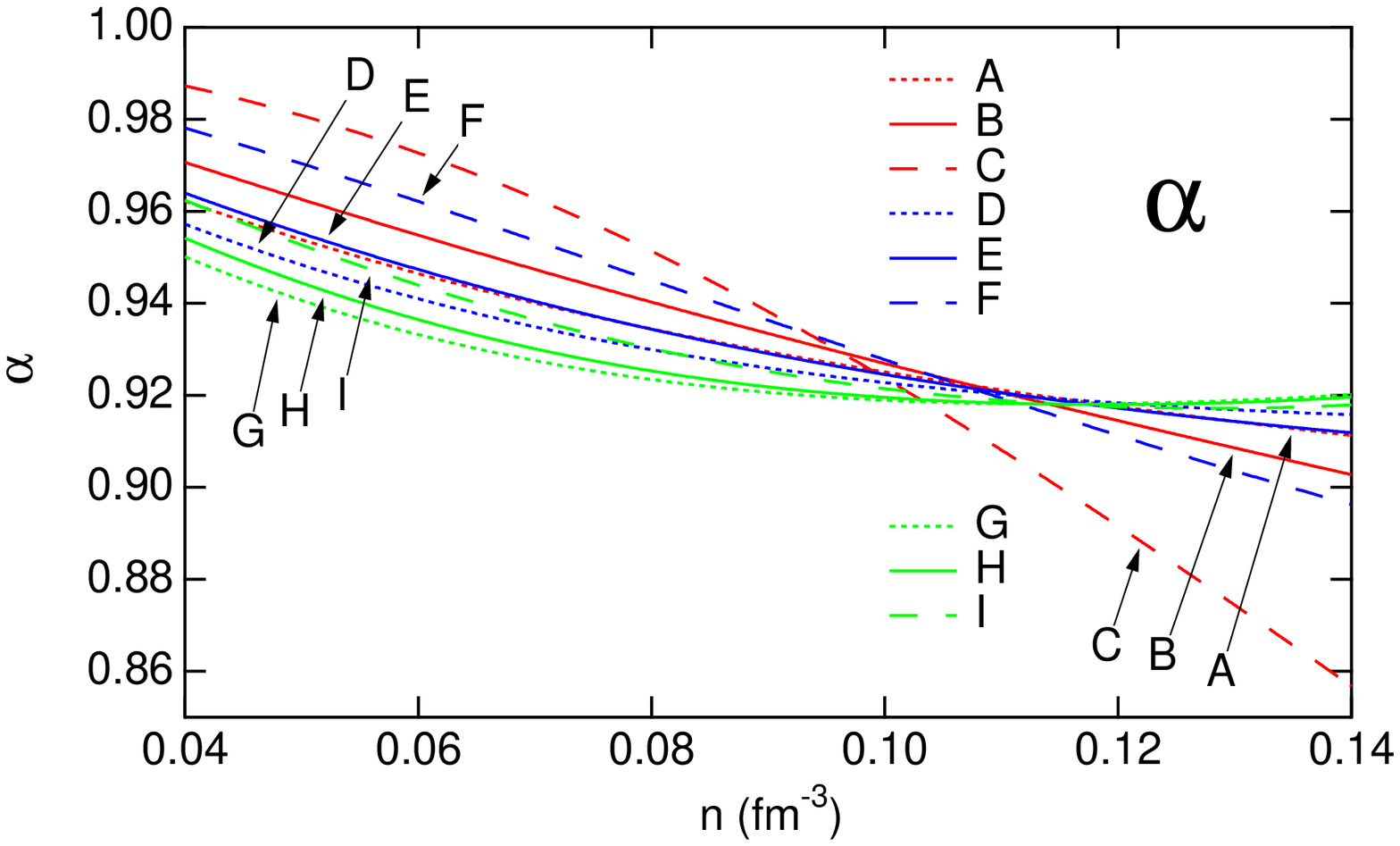}
\end{minipage}
\begin{minipage}{.45\linewidth}
 \includegraphics[width=8.5cm]{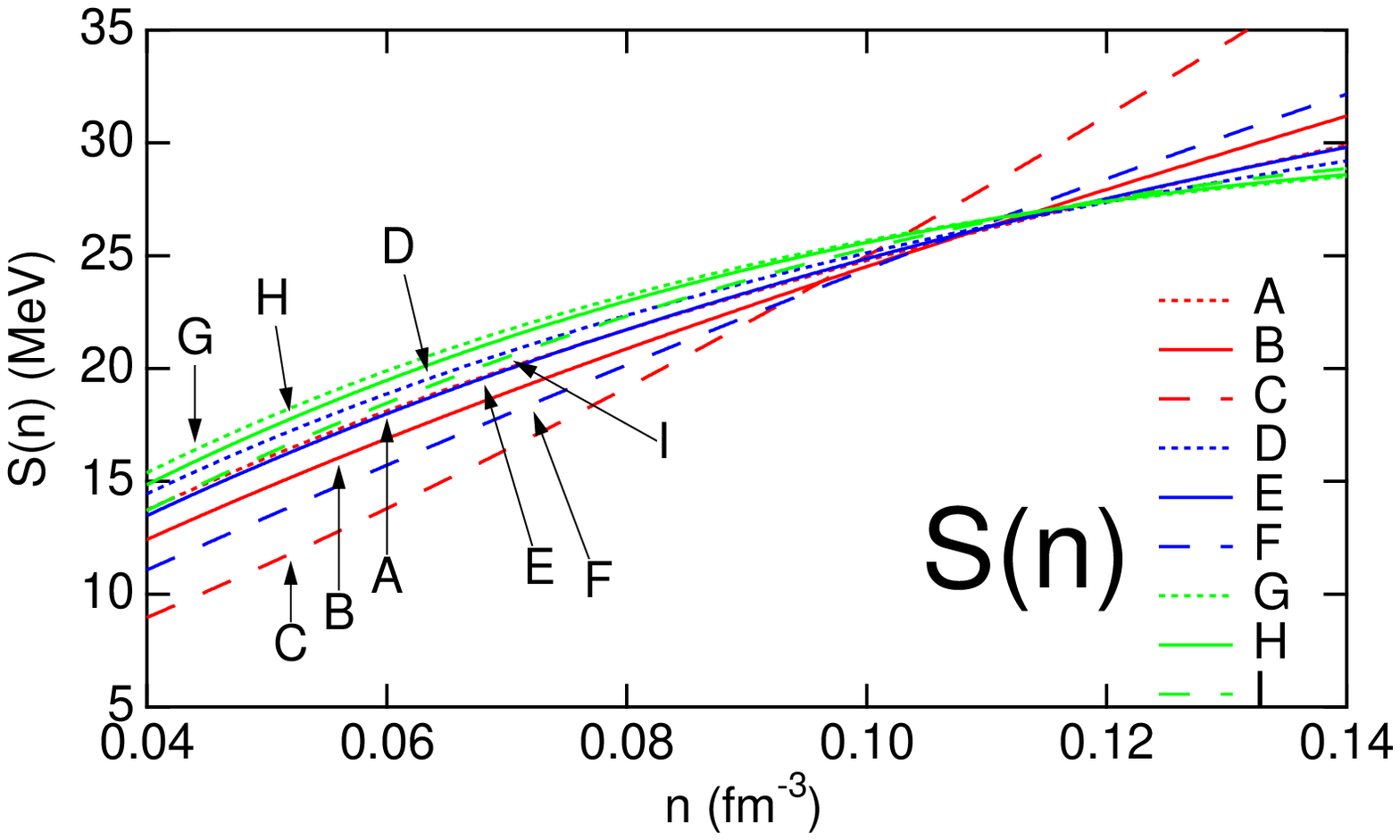}
\end{minipage}
\caption{\label{v} (Color online)
The proton effective potential $v_{\rm min}$, its bulk part
$v_0$, neutron excess $\alpha$, and symmetry energy coefficient $S(n)$ as a 
function of nucleon density, calculated for the EOS models A--I.}
\end{figure}

      In Fig.\ 8 we show the results for $n(Q)$ estimated for the parameter 
sets $(L,K_0)$ included in Fig.\ 1.  We find that $n(Q)$ decreases
with $L$, whereas it does not have a marked dependence on $K_0$.  
The $L$ dependence is correlated with the $L$ dependence of the symmetry 
energy coefficient $S(n)$ since $S(n)$ acts as a driving force of proton 
clustering.  Note the general tendency that
at subnuclear densities, the larger $L$, the smaller symmetry energy 
coefficient $S(n)$ (see Fig.\ 7).  The proton clustering thus takes place at 
lower density for larger $L$.  Figure 7 also shows that for $n\gtrsim0.1$ 
fm$^{-3}$, $S(n)$ becomes larger for larger $L$.  This is a feature coming 
from the empirical relation, $S_0\approx 0.075L+28$ MeV, derived in Ref.\ 
\cite{OI}.


\begin{figure}[t]
\begin{center}
\includegraphics[width=8.5cm]{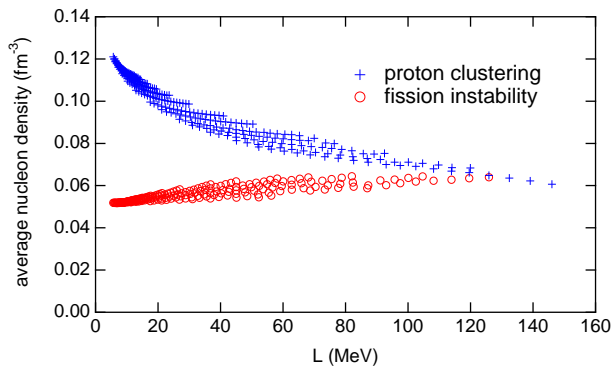}
\end{center}
\vspace{-0.5cm}
\caption{\label{nq} (Color online)
The onset density of proton clustering in uniform
nuclear matter as a function of $L$.  For comparison, we plot the density 
corresponding to $u=1/8$ in the phase with spherical nuclei, which is taken 
from Fig.\ 6(a).}
\end{figure}

      We can also observe from Fig.\ 8 that the difference
between $n(Q)$ and the density corresponding to $u=1/8$ in the phase with 
spherical nuclei  decreases with $L$ and eventually vanishes near $L=100$
MeV.  This suggests that the density regime of the pasta phases is limited
for a large $L$, although for the standard EOS model E, corresponding to
$L\simeq40$ MeV, it does appear between $\sim$0.06 and $\sim$0.09 fm$^{-3}$.
In our EOS model, a larger value of $L$ implies a harder
EOS of pure neutron matter as we shall see below.
We thus conclude that the absence of the pasta phases seen in Refs.\ 
\cite{LRP,Chen,SLy} from the EOS model with relatively high pressure of 
neutron matter at subnuclear densities is consistent with our result.

      In order to clarify this consistency, we calculate the pressure of 
pure neutron matter,
\begin{equation}
P_n=n_n^2 \left.\frac{\partial w}{\partial n_n}\right|_{\alpha=1},
  \label{Pn}
\end{equation}
for the parameter sets $(L,K_0)$ shown in Fig.\ 1.  The results for $P_n$ at 
$n_n=0.1$ fm$^{-3}$ are plotted as a function of $L$ in Fig.\ 9.  
We find out a roughly linear $L$ dependence of $P_n$ at $n_n=0.1$ fm$^{-3}$.
This dependence can be roughly understood by substituting the expansion 
(\ref{eos0}) into Eq.\ (\ref{Pn}) and thereby obtaining
\begin{equation}
P_n=\frac{K_0}{9}\left(\frac{n_n}{n_0}\right)^2(n_n-n_0)
     +\frac{L}{3}n_0 \left(\frac{n_n}{n_0}\right)^2.
  \label{Pn0}
\end{equation}
We remark that this pressure controls the neutron skin thickness of $^{208}$Pb 
evaluated within the framework of the Skyrme Hartree-Fock model \cite{Brown}.

\begin{figure}[t]
\begin{minipage}{.45\linewidth}
 \includegraphics[width=8.5cm]{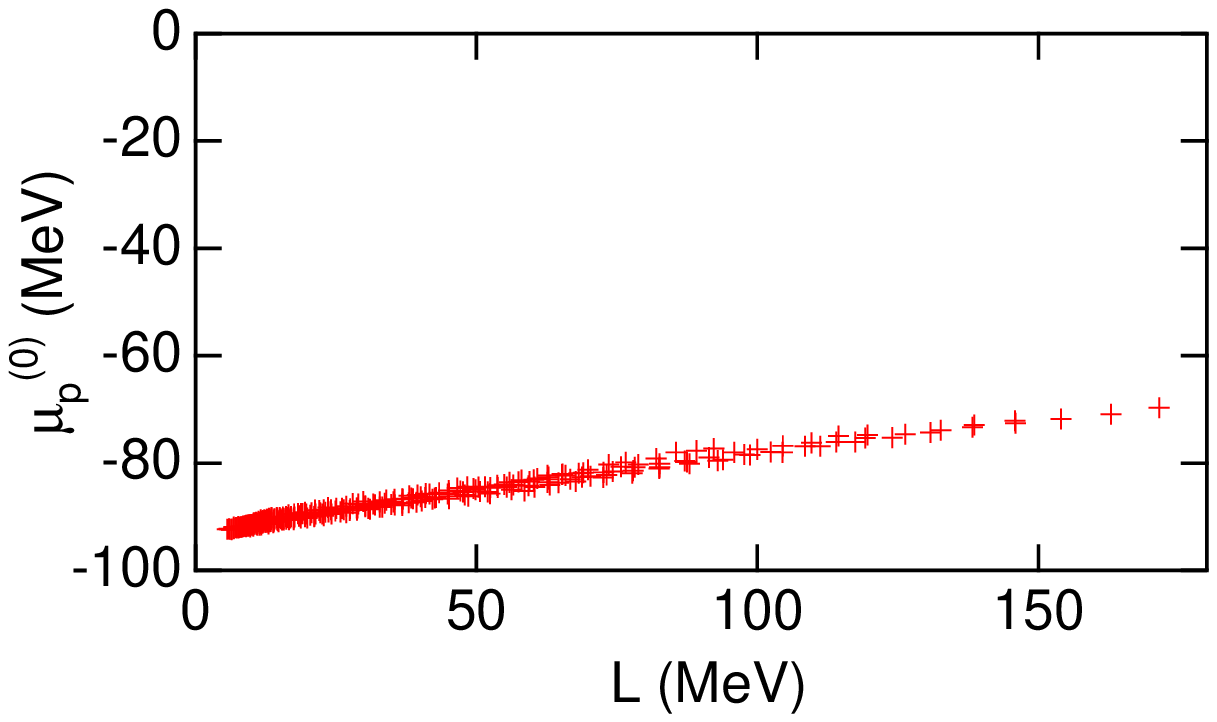}  
\end{minipage}
\begin{minipage}{.45\linewidth}
 \includegraphics[width=8.5cm]{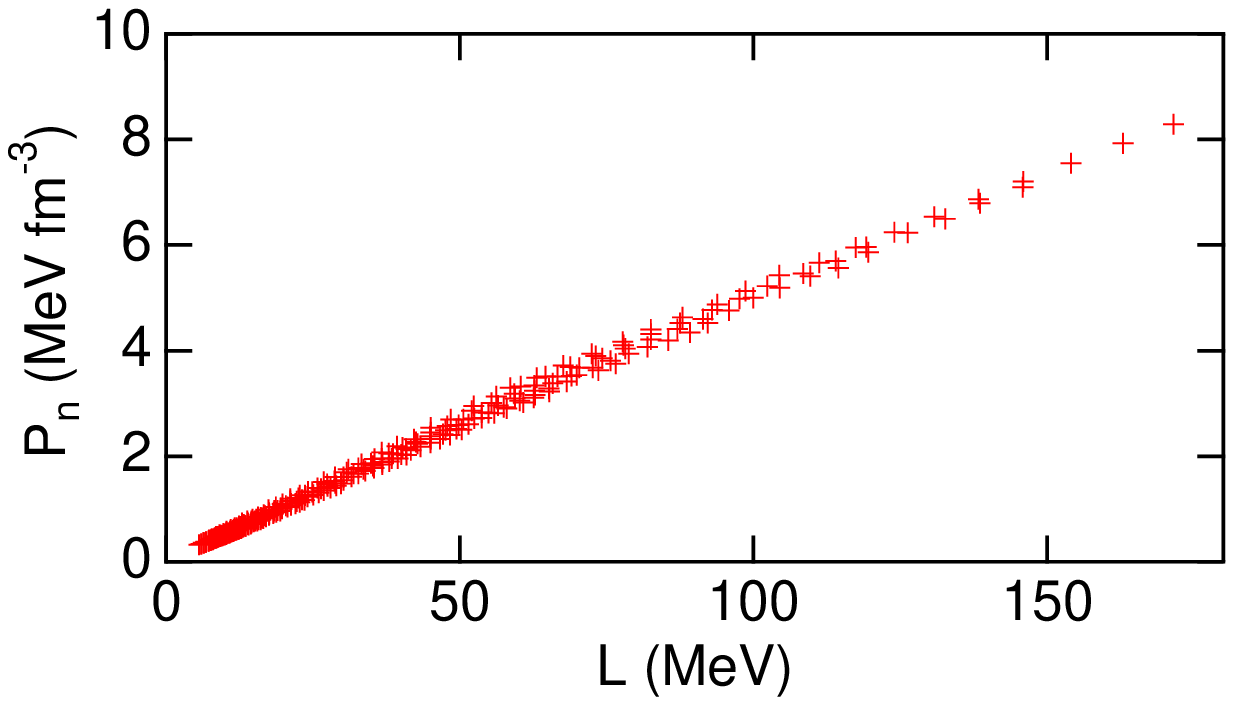}
\end{minipage}
\caption{\label{mup0pn}The proton chemical potential and pressure
in pure neutron matter of density 0.1 fm$^{-3}$ 
as a function of $L$.}
\end{figure}

      We conclude this section by mentioning a relation between the 
systematic liquid-drop analysis \cite{WIS} and the present analysis.
In Ref.\ \cite{WIS}, the value of $L$ was fixed at $L=60$ MeV, and 
the proton chemical potential in pure neutron matter, 
\begin{equation}
  \mu_p^{(0)}=\left.\frac{\partial(nw)}{\partial n_p}\right|_{\alpha=1},
   \label{mup0}
\end{equation}
was changed by a factor of 2.  In the present analysis, on the 
other hand, the value of $L$ was taken between 0 and 160 MeV, 
while $\mu_p^{(0)}$ depends only weakly on the value of $L$, 
as shown in Fig.\ 9 in which the results for $\mu_p^{(0)}$ calculated
at $n_n=0.1$ fm$^{-3}$ for the parameter sets $(L,K_0)$ shown in Fig.\ 1 
are plotted as a function of $L$.  According to Ref.\ \cite{WIS}, 
$\mu_p^{(0)}$ plays a role in shifting the density region of the pasta 
phases without changing its width significantly.  We may thus conclude 
that it is the parameter $L$ that controls the presence of the pasta phases.

\section{Conclusions}
\label{sec:conc}

     We have analyzed the equilibrium properties of inhomogeneous nuclear
matter at subnuclear densities in a way dependent on the density symmetry 
coefficient $L$ by using a macroscopic nuclear model.  We have estimated the 
upper and lower ends of the density region of the pasta phases from the 
onset densities of proton clustering in uniform nuclear matter and 
fission-like instability of spherical nuclei, respectively.  We find that
the upper end decreases with $L$, while the lower end is almost flat at
0.05--0.07 fm$^{-3}$.  The former arises from the $L$ dependence of the
symmetry energy, while the latter can be understood from the volume 
fraction $u\simeq1/8$ at which spherical nuclei become susceptible to
fission-inducing deformations.  For a typical EOS model consistent with the
GFMC calculations of pure neutron matter, the calculated pasta regime is 
appreciable.  In fact, the pasta regime is predicted to appear when
$L\lesssim100$ MeV.

     The present analysis is the first to attempt a systematic analysis of
the pasta region in terms of $L$.  However, much care needs to be taken of
the interpretation of the results.  While $L$ is the parameter 
characterizing the expansion of $w$ with respect to $n$ and $\alpha$ around 
$n=n_0$ and $\alpha=0$, the system of interest here is at large neutron excess
and at subnuclear densities.  The relation between the parameter $L$ and 
neutron star matter depends on the parametrization of $w$ with respect to $n$ 
and $\alpha$.  It is thus useful to keep in mind that we confined
ourselves to expressions (\ref{eos1})--(\ref{vn}) 
although they are known to be capable of 
reproducing various existing microscopic calculations of the EOS of uniform 
nuclear matter.

     We have also calculated the charge number $Z$ of spherical nuclei as 
a function of density for various values of $L$.  Generally, the charge number
$Z$ becomes smaller for larger $L$, a feature that could be of relevance to 
the evolution of neutron stars \cite{YP}.  In order to make better estimate 
of $Z$, however, shell and pairing effects should be taken into account.

\acknowledgments

      This work was supported in part by RIKEN through Grant No.\ A11-52040.


\end{document}